\documentstyle[11pt,epsfig]{article}
\title{\bf Accelerating universe in $f({\cal R})$ brane gravity}
\author{K. Atazadeh\thanks{email: k-atazadeh@sbu.ac.ir}\,,
M. Farhoudi\thanks{email: m-farhoudi@sbu.ac.ir}\,
and H. R. Sepangi\thanks{email: hr-sepangi@sbu.ac.ir}\\ {\small
Department of Physics, Shahid Beheshti University, Evin, Tehran
19839, Iran}}
\date{October 10, 2007}
\begin{document}
\maketitle

\begin{abstract}
We study a 5-dimensional $f({\cal R})$ brane gravity within the
framework of scalar-tensor type theories. We show that such a model
predicts, for a certain choice of $f({\cal R})$ and a spatially flat
universe, an exponential potential, leading to an accelerated
expanding universe driven solely by the curvature of the bulk space.
This result is consistent with the observational data in the
cosmological scale.
\end{abstract}
\vspace{2cm}
\section{Introduction}
The idea that our world might be a brane embedded in a higher
dimensional space-time (the bulk) \cite{1} has been in the
mainstream of cosmological investigations  in the past few years
\cite{3,4}. This approach differs from the usual Kaluza-Klein idea
in that the size of the extra dimensions can be large. The concept
of large extra dimensions is discussed phenomenologically in
\cite{5}. An important ingredient of the brane world scenario is
that the matter is confined to the brane and the only
communication between the brane and bulk is through gravitational
interaction or some other dilatonic matter. In general, the matter
on the brane leads to a cosmological evolution which is different
from the usual evolution governed by the Friedmann equation, that
is, in brane cosmology the Hubble parameter on the brane is
proportional to the square of energy density \cite{3,4}. This
proportionality is a result of the application of the Israel
matching condition which is basically a relation between the
extrinsic curvature and the energy-momentum tensor representing
matter fields on the brane.

Although in brane theories matter fields live on the brane, the
possibility of the presence of matter in the form of a scalar
field in the bulk has also been investigated in several works. One
of the first motivations to introduce a bulk scalar field was to
stabilize \cite{7} the distance between the two branes in the
context of the first model introduced by Randall and Sundrum
\cite{1}. A second motivation was the possibility of the
resolution of the famous cosmological constant problem \cite{8}.
Several works have studied, in particular, the impact of the
presence of a scalar field in the bulk on the cosmological
evolution on the brane, without trying to solve the full system of
equations in the bulk \cite{9,10}. In \cite{12}, the authors have
addressed some of the solutions for these equations and studied
the corresponding brane evolution. The purpose of the present
study is to employ modified gravity \cite{parry} in the Einstein
frame to explain the origin of such a self interacting scalar
potential.

An interesting observation made a few years ago was that the
expansion of our universe is currently undergoing a period of
acceleration which is directly measured from the light curves of
several hundred type Ia supernovae \cite{13} and independently from
observations of the cosmic microwave background (CMB) by the WMAP
satellite \cite{15} and other CMB experiments \cite{16}. However,
the mechanism responsible for this acceleration is not well
understood and many authors introduce a mysterious cosmic fluid, the
so called dark energy, to explain this effect \cite{17}. Recently,
it has been shown that such an accelerated expansion could be the
result of a modification to the Einstein-Hilbert action \cite{18} in
the framework of DGP brane cosmology. In the present work we study
the general form of the Einstein-Hilbert action for any function of
the Ricci scalar, $ f({\cal R})$, in  5 dimensions. This is done in
the framework of a scalar-tensor type theory \cite{20} where a
scalar field is minimally coupled to gravity with a self-interacting
potential. In this formulation we obtain explicit solutions using
conformal transformations, a technique employed in the case of an
empty bulk with a cosmological constant \cite{21} or a bulk with a
scalar field, similar to the present work, but with an exponential
potential. We present explicit solutions for a particular choice of
$ f({\cal R})$ which predict a similar exponential potential.

The organization of the manuscript is as follows: in section 2 we
briefly review the scalar-tensor formulation in 5-dimensions and
write the full system of equations. In section 3 we consider the
cosmological equations for $f({\cal R})$ gravity which, in the
Einstein frame, correspond to a self interacting scalar field with a
certain potential. Finally, we study the cosmological evolution on
the brane for ${\cal R}^{m}$ gravity which predicts a power law
acceleration in section 4. Conclusions are drawn in the last
section.
\section{Scalar-Tensor formulation of $f({\cal R})$ gravity}

Let us start from a general 5-dimensional action in the matter
frame
\begin{equation}\label{eq1}
{\cal S}[g_{_{AB}}]=\frac{1}{2\kappa_{5}^{2}}\int
d^{5}x\sqrt{-g}{f(\cal R)}+{\cal S}_{\rm m} [q_{\mu\nu},\psi_{\rm
m}].
\end{equation}
Here, $\kappa_{5}^{2}=8\pi G_{5}$, ${\cal R}$ is the 5-dimensional
scalar curvature and $f({\cal R})$ is some arbitrary function of the
scalar curvature with ${\cal S}\rm_{m}$ being the matter action
defined by the induced metric $q_{\mu\nu}$ and  the matter field
$\psi_{\rm m}$ on the brane. Under the conformal transformation
\cite{22}
\begin{equation}\label{eq2}
\tilde{g}_{_{AB}}=e^{\frac{\kappa_{5}}{\sqrt{3}}\Phi}g_{_{AB}},
\end{equation}
and the choice
\begin{equation}\label{eq3}
\Phi=\frac{2}{\sqrt{3}\kappa_{5}}\ln{f'({\cal R})},
\end{equation}
where the prime denotes derivative with respect to ${\cal R}$,
action (\ref{eq1}) can be written in the Einstein frame as \cite{20}
\begin{equation}\label{eq4}
\tilde{{\cal S}}[\tilde{g}_{_{AB}},\Phi]=\int d^{5}x
\sqrt{-\tilde{g}}\left[\frac{\tilde{{\cal R}}}
{2\kappa_{5}^{2}}-\frac{1}{2}(\tilde{\nabla}\Phi)^{2}-V(\Phi)\right]+\tilde{{\cal
S}}_{\rm m} [\tilde{q}_{\mu\nu},\psi_{\rm m}],
\end{equation}
where $\tilde{g}_{_{AB}}$ and
$\tilde{q}_{\mu\nu}=e^{\frac{1}{\sqrt{3}}\kappa_{5}\Phi}q_{\mu\nu}$
are the $5D$ bulk metric and the induced metric on the brane in the
Einstein frame respectively and $\tilde{{\cal R}}$ is the $5D$ Ricci
scalar associated with $\tilde{g}_{AB}$. One can show that the
effective potential in $5D$ is given by
\begin{equation}\label{eq5}
V(\Phi)=\frac{{\cal R}{f'({\cal R})}-{ f({\cal
R})}}{2\kappa_{5}^{2}{f'({\cal R})}^{5/3}}.
\end{equation}
This is the standard form of the scalar-tensor type theories
mentioned above. The $5D$ equations of motion corresponding to
action (\ref{eq4}) are
\begin{equation}\label{eq6}
\tilde{G}_{_{AB}}=\kappa_{5}^{2}\left[\tilde{T} _{_{AB}}(\psi_{\rm
m},\tilde{g})+{\tilde{\cal T}_{_{AB}}}(\Phi,\tilde{g})\right],
\end{equation}
where $\tilde{T}_{_{AB}}$ is the energy-momentum tensor in the
Einstein frame and ${\tilde{\cal T}}_{_{AB}}$ is given by
\begin{equation}\label{eq7}
\tilde{{\cal
T}}_{_{AB}}=\partial_{_{A}}\Phi\partial_{_{B}}\Phi-\tilde{g}_{_{AB}}
\left[\frac{1}{2}(\tilde{\nabla}_{_{C}}\Phi)(\tilde{\nabla}^{^{C}}\Phi)+
V(\Phi)\right],
\end{equation}
with the equation of motion for the scalar field as
\begin{equation}\label{eq8}
\tilde{\nabla}_{_{A}}\tilde{\nabla}^{^{A}}\Phi-\frac{dV(\Phi)}{d\Phi}=
\frac{\kappa_{5}}{2\sqrt{3}}e^{-\frac{2\kappa_{5}}{\sqrt{3}}\Phi}T\delta(y),
\end{equation}
where $T$ is the trace of energy momentum tensor in the Jordan
frame and $y$ represents the extra dimension. For cosmological
considerations, let us take a general form for the bulk metric in
the matter frame, also known as the Jordan frame, usually assumed
as
\begin{equation}\label{eq9}
ds^{2}=g_{_{AB}}dx^{A}dx^{B}=-n^{2}(y,t)dt^{2}+a^{2}(y,t)\gamma_{ij}dx^{i}dx^{j}+b^{2}(y,t)dy^{2},
\end{equation}
where $\gamma_{ij}$ is the maximally symmetric 3-dimensional
metric with $k=-1,0,1$ being the usual parameters denoting the
spatial curvature. Therefore, in the Einstein frame the metric is
$\tilde{g}_{_{AB}}$ and the functions $b(y,t)$, $a(y,t)$ and $dt$
can be written as
\begin{equation}\label{eq10}
\tilde{b}(y,\tilde{t})=e^{\frac{\kappa_{5}}{2\sqrt{3}}\Phi}b(y,t),
\end{equation}
\begin{equation}\label{eq11}
\tilde{a}(y,\tilde{t})=e^{\frac{\kappa_{5}}{2\sqrt{3}}\Phi}a(y,t)
\end{equation}
and
\begin{equation}\label{eq12}
d\tilde{t}=e^{\frac{\kappa_{5}}{2\sqrt{3}}\Phi}dt.
\end{equation}
Let us also take the matter on the brane as a perfect fluid, given
by
\begin{equation}\label{eq13}
\tilde{T}^{^{A}}\,_{_{B}}=\frac{1}{\tilde{b}(y,\tilde{t})}
\mbox{diag}\Big[\tilde{\rho}(\tilde{t}),\tilde{p}(\tilde{t}),\tilde{p}(\tilde{t}),\tilde{p}(\tilde{t}),0\Big]\delta(y),
\end{equation}
where
\begin{equation}\label{eq14}
\tilde{\rho}=e^{\frac{-2\kappa_{5}}{\sqrt{3}}\Phi}\rho,
\end{equation}
\begin{equation}\label{eq15}
\tilde{p}=e^{\frac{-2\kappa_{5}}{\sqrt{3}}\Phi}p.
\end{equation}
Here, $\tilde{\rho}$ and $\tilde{p}$ respectively are the energy
density and pressure in the Einstein frame. In the Gauss normal
coordinates,  $\tilde{g}_{_{55}}=\tilde{b}^{2}(y,\tilde{t})=1$,
the 5-dimensional bulk equations (\ref{eq6}) can be written as
\begin{equation}\label{eq16}
3\left\{\left(\frac{\dot{\tilde{a}}}{\tilde{a}}\right)^{2}-n^{2}\left[\frac{\tilde{a}''}{\tilde{a}}+
\left(\frac{\tilde{a}'}{\tilde{a}}\right)^{2}\right]+
k\frac{n^{2}}{\tilde{a}^{2}}\right\}=
\kappa_{5}^{2}\left[n^{2}V(\Phi)+\frac{1}{2}\dot{\Phi}^{2}+\frac{n^{2}}{2}\Phi'^{2}+n^{2}
\tilde{\rho}\delta(y)\right],
\end{equation}
\begin{equation}\label{eq17}
3\left(\frac{n'}{n}\frac{\dot{\tilde{a}}}{\tilde{a}}-\frac{\dot{\tilde{a}}'}{\tilde{a}}\right)=
\kappa_{5}^{2}\dot{\Phi}\Phi',
\end{equation}
\begin{eqnarray}\label{eq18}
3\left\{\frac{\tilde{a}'}{\tilde{a}}\left(\frac{\tilde{a}'}{\tilde{a}}+\frac{n'}{n}\right)-
\frac{1}{n^{2}}\left[\frac{\dot{\tilde{a}}}{\tilde{a}}\left(\frac{\dot{\tilde{a}}}{\tilde{a}}-
\frac{\dot{n}}{n}\right)+\frac{\ddot{\tilde{a}}}{\tilde{a}}\right]+\frac{k}{\tilde{a}^{2}}\right\}=
\kappa_{5}^{2}\left[V(\Phi)+\frac{1}{2n^{2}}\dot{\Phi}^{2}+\frac{1}{2}\Phi'^{2}\right]
\end{eqnarray}
and
\begin{eqnarray}\label{eq19}
\tilde{a}^{2}\left[\frac{\tilde{a}'}{\tilde{a}}\left(\frac{\tilde{a}'}{\tilde{a}}+2\frac{n'}{n}\right)+2\frac{\tilde{a}''}{\tilde{a}}+
\frac{n''}{n}\right]+\frac{\tilde{a}^{2}}{n^{2}}
\left[\frac{\dot{\tilde{a}}}{\tilde{a}}\left(-\frac{\dot{\tilde{a}}}{\tilde{a}}+2\frac{\dot{n}}{n}\right)-
2\frac{\ddot{\tilde{a}}}{\tilde{a}}\right]-k\\
\nonumber
=-\kappa_{5}^{2}\tilde{a}^{2}\left[\frac{1}{2}V(\Phi)-\frac{1}{2n^{2}}\dot{\Phi}^{2}+\Phi'^{2}-\tilde{p}\delta(y)\right].
\end{eqnarray}
The scalar field in the bulk, equation (\ref{eq8}), also reads
\begin{equation}\label{eq20}
\ddot{\Phi}+\left(3\frac{\dot{\tilde{a}}}{\tilde{a}}-\frac{\dot{n}}{n}\right)\dot{\Phi}-n^{2}\left[\Phi''+
\left(\frac{n'}{n}+3\frac{\tilde{a}'}{\tilde{a}}\right)\Phi'\right]+n^{2}\frac{dV(\Phi)}{d\Phi}
=n^{2}\frac{-\kappa_{5}}{2\sqrt{3}}e^{\frac{-2\kappa_{5}}{\sqrt{3}}\Phi}T\delta(y),
\end{equation}
where the prime and dot represent derivative with respect to $y$ and
$\tilde{t}$ respectively.

Assuming $\textbf{Z}_{2}$ symmetry and denoting
$\tilde{a}_{0}(\tilde{t})\equiv\tilde{a}(0^{+},\tilde{t})$,
$\tilde{a}'_{0}(\tilde{t})\equiv\tilde{a}'(0^{+},\tilde{t})$,
$n_{0}(\tilde{t})\equiv n(0^{+},\tilde{t})$,
$n'_{0}(\tilde{t})\equiv n'(0^{+},\tilde{t})$ and
$\Phi'_{0}(\tilde{t})\equiv\Phi'(0^{+},\tilde{t})$, we may proceed
to extract from the delta functions on both sides of the equations
(\ref{eq16}) and (\ref{eq19}), the matching conditions
\begin{equation}\label{eq21}
\frac{\tilde{a}'_{0}}{\tilde{a}_{0}}=-\frac{\kappa_{5}^{2}}{6}\tilde{\rho}(\tilde{t})
\end{equation}
and
\begin{equation}\label{eq22}
\frac{n'_{0}}{n_{0}}=\frac{\kappa_{5}^{2}}{6}\Big[3\tilde{p}(\tilde{t})+2\tilde{\rho}(\tilde{t})\Big].
\end{equation}
One notes that equation (\ref{eq21}) is consistent with the
assumption that the effect of extra dimension diminishes as one
moves away from the brane. Let us now turn to matching condition
for the scalar field. Using (\ref{eq20}) we obtain
\begin{equation}\label{eq23}
2\Phi'_{0}=\frac{\kappa_{5}}{2\sqrt{3}}e^{-\frac{2\kappa_{5}}{\sqrt{3}}\Phi_{0}}T^{^{\rm
(brane)}},
\end{equation}
where $T^{^{\rm (brane)}}=-\rho+3p$ is the trace for the
energy-momentum tensor in the matter frame. Application of the
matching condition for the scalar field leads to
\begin{equation}\label{eq24}
\Phi'_{0}=\frac{\kappa_{5}}{2\sqrt{3}}\gamma\tilde{\rho}(\tilde{t}),
\end{equation}
which involves all cases where the equation of state is of the
form $\tilde{p}=w\tilde{\rho}$ with $w$  as a constant and the
expression for $\gamma$ given by
\begin{equation}\label{eq25}
\gamma=\frac{1}{2}(3w-1).
\end{equation}
If the Lagrangian density for the perfect fluid is proportional to
the pressure as chosen in \cite{10}
\begin{equation}\label{eqq25}
{\cal L}(\phi)=-2{\cal F}(\phi)p(s,\varepsilon),
\end{equation}
where ${\cal F}(\phi)$ is an arbitrary function, $s$ and
$\varepsilon$ are entropy and enthalpy respectively, then in this
model we will have $\gamma=-4w\chi$. Note that in our model
$\chi=\frac{1}{2}$. Thus, if the matter content of the brane
behaves like a cosmological constant, $w=-1$, this model will be
compatible with that presented in \cite{10}.

\section{Cosmological equations on the brane}
In this section we consider the cosmological behavior on the brane
using the global equations obtained in the previous section. For
the brane, assumed to stay at $y=0$, the Einstein frame induced
FRW metric with $k=0$ is
\begin{equation}\label{eq26}
d\tilde{s}^{2}=-n^{2}_{0}(\tilde{t})d\tilde{t}\,^{2}+\tilde{a}^{2}_{0}(\tilde{t}\,)\delta_{ij}dx^{i}dx^{j}.
\end{equation}
One may now proceed to obtain the cosmological equations by taking
the gauge
\begin{equation}\label{eq27}
n_{0}(\tilde{t})=1.
\end{equation}
The cosmic time $\tau$ in the Einstein frame can be derived from the
$\tilde{t}$ by
\begin{equation}\label{eq28}
\tau=\int^{\tilde{t}}n_{0}(t')dt'.
\end{equation}
This gauge is convenient because it gives the usual cosmological
time on the brane.

Let us now obtain the Friedmann equation as well as a generalized
conservation equation on the brane. We will closely follow the
derivation presented in \cite{4,12} with the additional ingredient
of an energy flux from the fifth dimension, {\it i.e.} the
component (0,5) of the bulk energy-momentum tensor is assumed to
be non zero because of the presence of the scalar field. Use of
the matching conditions in the (0,5) component of the field
equations, (\ref{eq17}), in the Einstein frame evaluated on the
brane yields the generalized conservation equation
\begin{equation}\label{eq29}
\dot{\tilde{\rho}}+3\frac{\dot{\tilde{a}}_{0}}{\tilde{a}_{0}}(\tilde{\rho}+\tilde{p})=2\tilde{{\cal
T}}_{05}\Big|_{_{y=0}},
\end{equation}
where $\tilde{{\cal T}}_{05}=\dot{\Phi}\Phi'$. Now, by using the
matching condition (\ref{eq24}) for the scalar field, it reads
\begin{equation}\label{eq30}
\dot{\tilde{\rho}}+3\frac{\dot{\tilde{a}}_{0}}{\tilde{a}_{0}}(\tilde{\rho}+\tilde{p})=
\gamma\dot{\bar{\Phi}}_{0}\tilde{\rho},
\end{equation}
where $ \bar{\Phi}\equiv\frac{\kappa_{5}}{\sqrt{3}}\Phi$ and the dot
represents derivative with respect to $\tau$. This equation is the
generalized conservation law for cosmological matter. For an
equation of state $\tilde{p}=w\tilde{\rho}$ with $w$ constant, the
integration of equation (\ref{eq30}) yields the following evolution
for the energy density
\begin{equation}\label{eq31}
\tilde{\rho}\propto\tilde{a}_{0}^{-3(1+w)}e^{\gamma\bar{\Phi}_{0}}.
\end{equation}
If the scalar field is constant in time we will recover the familiar
evolution of the standard cosmology. Let us now consider the (5,5)
component of the field equations, (\ref{eq18}). Using the (0,5)
component, it can be rewritten in the form
\begin{equation}\label{eq32}
\dot{F}=\frac{2}{3}\dot{\tilde{a}}_{0}\tilde{a}_{0}^{3}\kappa_{5}^{2}\tilde{{\cal
T}}^{5}\,_{5}\Big|_{_{y=0}}-\frac{2}{3}\tilde{a}'_{0}\tilde{a}_{0}^{3}\kappa_{5}^{2}\tilde{{\cal
T}}^{5}\,_{0}\Big|_{_{y=0}},
\end{equation}
with
\begin{equation}\label{eq33}
F\equiv
(\tilde{a}_{0}\tilde{a}'_{0})^{2}-(\tilde{a}_{0}\dot{\tilde{a}}_{0})^{2}.
\end{equation}
This corresponds to a slight generalization of the expression given
in \cite{4}. The expression for the (5,5) component of the
energy-momentum tensor of the scalar field  $\tilde{{\cal
T}}_{_{AB}}$ is given by
\begin{equation}\label{eq36}
\tilde{{\cal
T}}^{5}\,_{5}=\frac{1}{2}\left(\Phi'^{2}+\dot{\Phi}^{2}\right)-V(\Phi).
\end{equation}
Using equation (\ref{eq32}) and the matching conditions, one
obtains, after integrating  the time, the following generalized
Friedmann equation in the Einstein frame
\begin{equation}\label{eq34}
\tilde{H}^{2}_{_{0}}=\frac{\kappa_{5}^{4}}{36}\tilde{\rho}^{2}-\frac{2\kappa_{5}^{2}}{3\tilde{a}_{0}^{4}}\int
d\tau\dot{\tilde{a}}_{0}\tilde{a}_{0}^{3}\tilde{{\cal
T}}^{5}\,_{5}\Big|_{_{y=0}}-\frac{\kappa_{5}^{4}\gamma}{18\tilde{a}_{0}^{4}}\int
d\tau\dot{\bar{\Phi}}_{0}\tilde{a}_{0}^{4}\tilde{\rho}^{2},
\end{equation}
where the Hubble parameter is defined by
\begin{equation}\label{eq35}
\tilde{H}_{_{0}}\equiv\frac{\dot{\tilde{a}}_{0}}{\tilde{a}_{0}},
\end{equation}
and the constant of integration is taken to be zero. The quadratic
appearance of the energy density in this equation is a generic
feature of the brane cosmology \cite{3}. It also has an integral
term related to the pressure along the fifth dimension and an
integral term related to the energy flux coming from the bulk
scalar field.

Now, using the matching condition (\ref{eq24}) we obtain
\begin{equation}\label{eq37}
\frac{2}{3}\kappa_{5}^{2}\tilde{{\cal T}}^{5}\,_{5}
\Big|_{_{y=0}}=\frac{\kappa_{5}^{4}}{36}\gamma^{2}\tilde{\rho}^{2}+\dot{\bar{\Phi}}^{2}_{0}-
\frac{2}{3}\kappa_{5}^{2}V(\Phi)\Big|_{_{y=0}}.
\end{equation}
Finally, after evaluating (\ref{eq20}) at $y=0$ together with the
use of the matching conditions (\ref{eq21}), (\ref{eq22}) and
(\ref{eq24}) the scalar field equation on the brane is given by
\begin{equation}\label{eq38}
\ddot{\Phi}_{0}+3\left(\frac{\dot{\tilde{a}}_{0}}{\tilde{a}_{0}}\right)\dot{\Phi}_{0}-\hat{\Phi}''_{0}+\frac{dV(\Phi)}{d\Phi}
\Big|_{_{y=0}}
=\frac{\kappa^{3}_{5}}{6\sqrt{3}}\gamma^{2}\tilde{\rho}^{2}(\tau),
\end{equation}
where $\hat{\Phi}''_{0}$ stands for the non-distributional part of
the scalar field derivative. Thus equations (\ref{eq34}) and
(\ref{eq38}) are the equations of motion for the evolution of the
cosmic on the brane in the Einstein frame. In the next section we
will examine these equations for a particular choice of $f(R)$
gravity.
\section{Cosmological evolution in ${\cal R}^{m}$ gravity}

We start from the Lagrangian
\begin{equation}\label{eq39}
f({\cal R})= f_{0}{\cal R}^{m},
\end{equation}
for which potential (\ref{eq5}) is given by
\begin{equation}\label{eq40}
V(\Phi)= V_{0}e^{\alpha\bar{\Phi}},
\end{equation}
where
\begin{equation}\label{eq41}
V_{0}=\frac{1}{2\kappa_{5}^{2}}f_{0}(m-1)(mf_{0})^{\frac{m}{1-m}},
\end{equation}
with $\alpha\equiv\frac{-2m+5}{2(m-1)}$ and $f_{0}$ is a constant.

As it can be seen, the exponent in the above potential is singular
for $m=1$ and therefore warrants further discussion. For this value
of $m$, the scalar field $\Phi$ from equation (\ref{eq3}) becomes
constant and we have $\tilde{g}_{_{AB}}=\mbox{const.}\times
g_{_{AB}}$, indicating that the Jordan frame is equivalent to the
Einstein frame. Also, the effective potential for $m=1$ in the
Einstein frame is zero, similar to what one obtains in an empty $4D$
universe for which the dynamics is governed by the same Lagrangian
\cite{20,22}. This seems to be a general feature of modified
theories of gravity when the Lagrangian is of the form (\ref{eq39}).
In what follows, we determine the range of validity for $m$ which
would allow the universe to achieve an accelerated expansion.

Now, we assume that both the total energy density $\rho$ and
pressure $p$ on the brane consist of two parts
\begin{equation}\label{eqq42}
\rho=\lambda+\varrho~~~~~~~~\mbox{and}~~~~~~~~~p=-\lambda+\textsf{p},
\end{equation}
where $\lambda$, $\varrho$ and $\textsf{p}$ are the tension, the
usual cosmological energy density and pressure in the matter
frame, respectively. In what follows we concentrate on the case
$\varrho=\textsf{p}=0$, {\it i.e.} the vacuum solution. Equation
(\ref{eq25}) then implies that $\gamma=-2$. One notes that by
retaining a non-zero effective tension on the brane we are
actually taking the brane effects into account. For simplicity and
following \cite{sasaki}, we take the tension, $\lambda$, in the
matter frame as
\begin{equation}\label{tension}
\lambda=\bar{\lambda}_{c}e^{(\frac{\alpha}{2}+2)\bar{\Phi}},
\end{equation}
where $\bar{\lambda}_{c}\equiv\frac{\lambda_{c}}{\kappa_{5}^{2}}$
and $\lambda_{c}$ is a constant. Therefore,
$\tilde{\lambda}=e^{-2\bar{\Phi}}\lambda=\bar{\lambda}_{c}e^{\frac{\alpha}{2}\bar{\Phi}}$
is the brane tension in the Einstein frame. Thus, the equations of
motion on the brane, (\ref{eq34}) and (\ref{eq38}), become
\begin{equation}\label{eq42}
\tilde{H}^{2}_{_{0}}=\frac{\kappa^{4}_{5}}{36}\tilde{\lambda}^{2}-\frac{2\kappa_{5}^{2}}{3\tilde{a}_{0}^{4}}\int
d\tau\dot{\tilde{a}}_{0}\tilde{a}_{0}^{3}\tilde{{\cal
T}}^{5}\,_{5}\Big|_{_{y=0}}-\frac{\kappa^{4}_{5}}{9\tilde{a}_{0}^{4}}\int
d\tau\tilde{\lambda}\tilde{a}_{0}^{4}\tilde{{\cal
T}}^{5}\,_{0}\Big|_{_{y=0}}
\end{equation}
and
\begin{equation}\label{eq43}
\ddot{\Phi}_{_{0}}+3\left(\frac{\dot{\tilde{a}}_{0}}{\tilde{a}_{0}}\right)\dot{\Phi}_{_{0}}-\hat{\Phi}''_{0}+
\frac{dV(\Phi)}{d\Phi}\Big|_{{_{y=0}}}=\frac{2\kappa^{3}_{5}}{3\sqrt{3}}\tilde{\lambda}^{2}.
\end{equation}
These equations are the basic equations of motion on the brane
without matter in the Einstein frame. We now look for a power law
solution for the scale factor. Substituting the ans\"{a}tze
\begin{equation}\label{eq44}
\tilde{a}_{0}(\tau)\propto\tau^{\beta}~~~~~~~~\mbox{and}~~~~~~~~~\Phi_{0}(\tau)=\sigma\ln\tau
\end{equation}
into equations (\ref{eq42}) and (\ref{eq43}), we find
\begin{equation}\label{eq45}
\sigma=-2\frac{\sqrt{3}}{\kappa_{5}\alpha},
\end{equation}
where $\tau\neq0$ and $\alpha\neq0$, {\it i.e.} $m\neq5/2$. Now,
using the above value for $\sigma$ into equations (\ref{eq42}) and
(\ref{eq43}) we have
\begin{equation}\label{eqq46}
(4\beta-2)\beta^2-\left(\frac{4\beta}{9}-\frac{2}{9\alpha}-\frac{1}{18}\right)\lambda_c^2+\frac{4
\left(2 \beta -3 \beta^2\right)}{\alpha ^2}=0.
\end{equation}
This algebraic equation has one explicit real solution for $\beta$
in terms of $\alpha$ and $\lambda_{c}$. To obtain the functional
dependence of $\tau$, we note that it is the cosmic time in the
Einstein frame which is related to coordinate $t$ in the matter
frame by $e^{-\frac{1}{2}\bar{\Phi}_{0}}d\tau=dt$. As a result
\begin{equation}\label{eq47}
\tau=\left(\frac{\alpha+1}{\alpha}\right)^{\frac{\alpha}{\alpha+1}}t^{\frac{\alpha}{\alpha+1}},
\end{equation}
up to a constant of integration, noting that $\alpha$ cannot take
the value $-1$ by definition. The scale factor in the physical
(Jordan or matter) frame is thus given by
\begin{equation}\label{eq48}
a_{0}(t)=e^{-\frac{1}{2}\bar{\Phi}_{0}}\tilde{a}_{0}(\tau)\propto\left(\frac{\alpha+1}{\alpha}\right)^
{\frac{\alpha\beta+1}{\alpha+1}}t^{\frac{\alpha\beta+1}{\alpha+1}}.
\end{equation}
Equation (\ref{eq48}) shows that there is a possibility of having an
accelerated expanding universe for some choices of $m$ and
$\lambda_{c}$.

The deceleration parameter on the brane as a function of $m$ and
$\lambda_{c}$ is therefore given by
\begin{equation}\label{eq49}
q(m,\lambda_{c})=-\frac{a_{0}\ddot{a}_{0}}{\dot{a}_{0}^{2}}=-\frac{\alpha\beta-\alpha}{\alpha\beta+1}.
\end{equation}
The condition for acceleration, $q(m,\lambda_{c})<0$, in equation
(\ref{eq49}) leads to $\beta>1$ from which, using definition
$w_{\rm eff}=-1-\frac{2\dot{H}_{_{0}}}{3H_{_{0}}^2}$ for the
effective quintessence, we find $w_{\rm eff}<-1/3$. Figure 1 shows
the behavior of the deceleration parameter, $q$, as a function of
$m$ and $\lambda_{c}$. As it can be seen, for
$m\longrightarrow-\infty$ and $\lambda_{c}\rightarrow\pm\infty$ we
have $q\rightarrow-1$, that is the universe finally approaches the
eternal de Sitter phase. The range of validity of $m$ shown in
figure 1 is consistent with the observational SNeIa data in
$4$-dimensional $f({\cal R})$--models \cite{capo}. It is therefore
plausible that modified gravity within the context of brane
theories presents an alternative to dark energy with the
possibility of having an accelerated expanding universe.

A point worth emphasizing again is that, the universe in our
model, taken to be devoid of ordinary matter,  would undergo an
accelerated expansion for all values of $\lambda_{c}$ if the value
for $m$ is within the range shown in figure 1 which excludes the
value $m=1$ as well. For this value of $m$, the two frames, namely
the Jordan and Einstein frames coincide and $V(\Phi)=0$. As was
mentioned above, the same behavior is also manifest in
$4$-dimensional $f({\cal R})$--models where the universe is taken
to be empty \cite{20,22}. This points to a typical behavior in
$f({\cal R})\sim{\cal R}^{m}$ theories, both in four and five
dimensions, where for $m=1$ the resulting universe in the present
context is a static one.
\begin{figure}
\begin{center}
\epsfig{figure=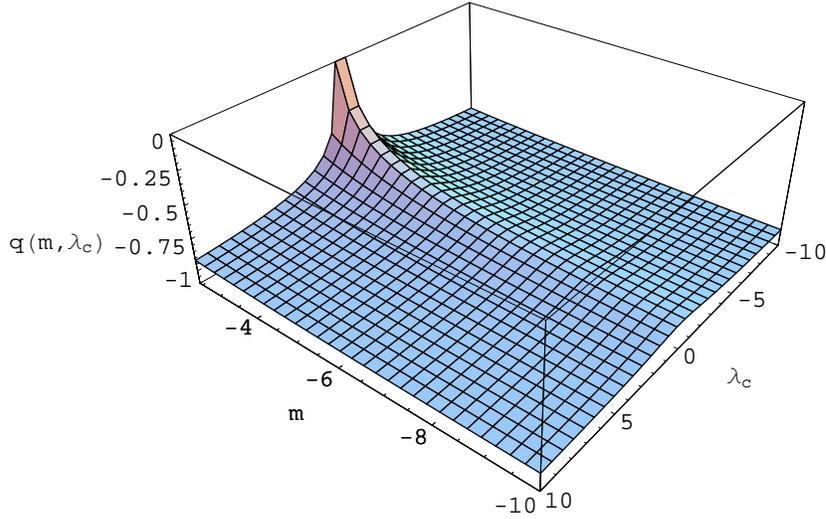,width=11cm}\hspace{5mm}
\end{center}
\caption{\footnotesize Behavior of $q(m,\lambda_{c})$ as a function
of $m$ and $\lambda_{c}$. An accelerating universe occurs for
$m\leq-2.42$ and for all values of $\lambda_{c}$.}
\end{figure}
\section{Conclusions}
In this manuscript we have obtained explicit solutions in a brane
world scenario where an arbitrary function of the Ricci scalar is
taken as the bulk Lagrangian. Using a conformal transformation,
the action is converted to that of a scalar-tensor type theory
with a scalar field. We have shown that with a suitable choice for
the function $f({\cal R})$ and brane tension $\lambda$, an
accelerated expanding universe emerges. The source of this
acceleration is not related to an exotic matter but to a scalar
field whose origin can be traced back to geometry of the brane
and, specifically, to the curvature scalar $\cal R$ and depends on
two free parameters, namely $\alpha$ and $\lambda_c$. Hence, an
accelerating universe driven by curvature would certainly seem to
be a possibility.

\end{document}